\documentclass[twocolumn, prb, showpacs, amsmath, amssymb, superscriptaddress]{revtex4}

\usepackage{graphicx}
\usepackage{dcolumn}
\usepackage{bm}

\begin{document}

\title{Dynamic critical behaviors in two-dimensional Josephson junction 
arrays with positional disorder}

\author{Jaegon Um}
\affiliation{National Creative Research Initiative Center for Superconductivity, POSTECH, 
Pohang 790-784, Korea}
\author{Beom Jun Kim}%
\email[Corresponding author; ]{beomjun@skku.edu}
\affiliation{Department of Physics, BK21 Physics Research Division,  and Institute of Basic Science, Sungkyunkwan University, Suwon 440-746, Korea}
\author{Petter Minnhagen}
\affiliation{Department of Physics, Ume{\aa} University,
90187 Ume{\aa}, Sweden}
\author{M.Y. Choi}
\affiliation{Department of Physics and Center for Theoretical Physics, Seoul National University, Seoul 151-747, Korea}
\affiliation{Korea Institute for Advanced Study, Seoul 132-722, Korea}
\author{Sung-Ik Lee}
\affiliation{National Creative Research Initiative Center for Superconductivity, POSTECH,
Pohang 790-784, Korea}
\affiliation{Quantum Material Laboratory, Korea Basic Science Institute, Daejon 305-333, Korea}

\date{\today}

\begin{abstract}
We numerically investigate dynamic critical behaviors of two-dimensional (2D) 
Josephson-junction arrays with positional disorder in the scheme of 
the resistively shunted junction dynamics. 
Large-scale computation of the current voltage characteristics reveals
an evidence supporting that a phase transition occurs at a nonzero critical 
temperature in the strong disorder regime, as well as in the weak disorder regime. 
The phase transition at weak disorder appears to belong to 
the Berezinskii-Kosterlitz-Thouless (BKT) type. In contrast, evidence
for a non-BKT transition is found in the strong disorder regime.
These results are consistent with the recent experiment 
on positionally disordered Josephson-junction arrays;
in particular, the critical temperature of the non-BKT transition 
(ranging from 0.265 down to the minimum 0.22 in units of $E_J/k_B$ 
with the Josephson coupling strength $E_J$), 
the correlation length critical exponent $\nu=1.2$,
and the dynamic critical exponent $z=2.0$ in the strong disorder regime 
agree with the existing studies of the 2D gauge-glass model. 
\end{abstract}


\maketitle

\section{\label{sec:intro} Introduction}

Phase transitions for the two-dimensional (2D) Josephson-junction arrays (JJAs)
with weak positional disorder (i.e., on a slightly disordered lattice) and in a
perpendicular magnetic field have attracted much attention.  
Here positional disorder in the presence of an external magnetic field
effectively induces random phase shifts of the magnetic bond 
angles;~\cite{forrester,kosterlitz}
thus the corresponding 2D random gauge $XY$ (RG$XY$) model,  in which
the magnetic bond angles are quenched random variables distributed
in a certain width,  provides a 
theoretical realization of such a positionally disordered Josephson-junction array (PDJJA) 
in a magnetic field:
Strong magnetic field in the PDJJA corresponds to strong
disorder in the RGXY model.  
It has been observed that sufficiently weak disorder in the RG$XY$ model does not destroy 
the quasi-long-range order present at low temperatures and accordingly, the system
undergoes a Berezinskii-Kosterlitz-Thouless (BKT) type transition 
at a finite critical temperature $T_{c}$, 
which decreases as the disorder strength is raised up to a critical value.~\cite{kosterlitz, akino, holme}
On the other hand, in spite of a number of studies, the strong disorder regime 
of the RG$XY$ model (and its fully disordered limit corresponding to the gauge-glass model)
has resisted adequate understanding. 
In parallel with the Mermin-Wagner theorem~\cite{mermin} for the absence of
the long-range order in the $XY$ model, Nishimori~\cite{nishimori} has proven
the absence of the long-range glass order in the gauge-glass model. However, 
it should be noted that the vanishing local glass order parameter in 
the gauge glass is not completely incompatible with the existence of 
the superconducting phase at nonzero temperatures. For instance, 
quasi-long-range glass order~\cite{choi} and possibility of
a continuous phase transition of an anomalous dimension~\cite{holme}
have been suggested. 

On the one hand, there exist numerical evidences supporting
the zero-temperature phase transition:
Domain-wall renormalization-group studies~\cite{kosterlitz, akino} predict that
the RG$XY$ model in the strong disorder regime as well as the gauge-glass model
undergoes a phase transition only at zero temperature. 
For the gauge-glass model, the zero-temperature transition was
supported by computations of various quantities such as
the current-voltage ($IV$) characteristics,~\cite{hyman, granato}
the root-mean-square current,~\cite{young, nikolaou} 
the correlation length,~\cite{katzgraber, katzgraber2}
the glass susceptibility,~\cite{young, katzgraber, katzgraber2}  
the autocorrelation function,~\cite{katzgraber2, nikolaou}
and the phase slip resistance.~\cite{nikolaou}
A recent numerical consideration of the low-energy excitations 
also estimated $T_c =0$ in the gauge-glass model.~\cite{tang}      
   
In contrast, 
the finite-size scaling analysis applied to the helicity modulus and the root-mean-square current 
in the RG$XY$ model with strong disorder 
demonstrated that the superconducting phase transition of a non-BKT type
occurs at finite $T_c$, independent of the disorder strength.~\cite{holme} 
In fact evidences for such a finite-temperature transition 
in the gauge-glass model have been presented in a few numerical studies 
computing the $IV$ characteristics,~\cite{li, chen}
the correlation function,~\cite{choi} the glass susceptibility,~\cite{choi}
and the linear resistance.~\cite{kim} 
Here it should be noted that critical exponents and $T_{c}$ obtained from the resistively
shunted junction (RSJ) simulations in large scales~\cite{chen} 
agree with other numerical studies of the gauge glass,~\cite{choi, kim}
and lead to $T_{c}=0.22$ (in units of $E_J/k_{B}$, where $E_J$ is the
coupling energy and $k_{B}$ the Boltzmann constant), the correlation length
critical exponent $\nu = 1.2$, and the dynamic critical exponent $z=2.0$. 
Furthermore, the barrier energy and the associated vortex mobility 
in the gauge glass~\cite{holme2} implies that superconducting order 
persists at low but finite temperatures.  

Very recently, an experiment has been performed on the PDJJA,~\cite{yun}  
and the critical temperature has been measured as a function of the disorder strength, 
which reveals a finite-temperature transition at strong disorder.
Further, in the experiment~\cite{yun} the scaling behavior of the $IV$ characteristics 
is observed consistent with the corresponding numerical results for the gauge glass,\cite{li,chen} 
indicating the presence of a non-BKT type transition at a finite temperature 
in the strong disorder regime. 
%
%
A truly remarkable feature of this PDJJA experiment 
is that a very clear signal of a finite-temperature transition has been obtained for a 
{\em single} disorder realization of a $200\times 800$ square array. 
In contrast, typical attempts to settle the issue of the existence of a finite-temperature transition 
by direct simulations usually presume that a well converged disorder average is necessary. 
This in turn limits the simulations to small system sizes, so that any conclusion crucially
depends on {\it a priori} assumptions for the finite-size scaling. 
The PDJJA experiment in Ref.~\onlinecite{yun} suggests that one could 
avoid such {\it a priori} assumptions by instead obtaining well 
converged data for large samples at the price of focussing on a single disorder realization. 

In the present paper we test this alternative 
strategy by performing RSJ simulations and computing the $IV$-characteristics
of a large square array of Josephson junctions with one randomly chosen positional disorder realization. 
In parallel with the experimental results in Ref.~\onlinecite{yun}, we find strong
numerical evidence for a finite-temperature transition.  In view of the on-going
controversy as to the existence of such a phase transition, we discuss the
possible implications of our results.

The present paper is organized as follows: In Sec.~\ref{sec:model}, we briefly
describe the numerical method used in this work, and discuss the relation
between the PDJJA and the RG$XY$ model. Section~\ref{sec:result1} presents mostly 
the results for the PDJJA at weak disorder, which is followed by
the results at strong disorder in Sec.~\ref{sec:result2}. 
In Sec.~\ref{sec:summary}, we summarize the results of the present paper in relation with
the existing studies. 

\section{\label{sec:model}
Model and Simulations}
With the assumption that the coupling energy is not affected by positional
disorder, the Hamiltonian of the PDJJA in the presence of 
a transverse magnetic field reads~\cite{footnote}
\begin{equation}
H = -E_J \sum_{\langle ij\rangle} \cos \left( \phi_{ij} 
- A_{ij} \right),
\label{eq:H}
\end{equation}  
where $\phi_{ij}$ is the phase difference between the superconducting
islands at sites $i$ and $j$.
When an external magnetic field $\mathbf{B}= B \hat{\bf z}$ is applied,
the magnetic bond angle $A_{ij}$ obtains the form 
\begin{equation}
\label{gauge}
A_{ij} = \frac{B a^{2} \pi}{\Phi_{0}} \left( x_{j}+x_{i} \right)
\left( y_{j}-y_{i} \right),
\end{equation}
where $a$ is the lattice constant and $\Phi_{0}$ is the magnetic 
flux quantum. Due to the positional disorder, the position of the
$i$th island is given by
\begin{equation}
\mathbf{r}_{i} \equiv\left( x_{i} , y_{i} \right) = \left( x_{i}^{0}{+} \delta x_{i} , y_{i}^{0}{+} \delta y_{i} \right),
\label{positional_disorder}
\end{equation}
where $\mathbf{r}_{i}^{0} \equiv \left( x_{i}^{0} , y_{i}^{0} \right)$
represents the ideal position without  disorder 
and $\delta x_{i}$ and $\delta y_{i}$ are random quenched variables uniformly distributed in 
the interval $[ - \Delta, \Delta ]$. Henceforth, $\mathbf{r}_i$,
$x_i$, $y_i$, $\delta x_i$, and $\delta y_i$ are all taken to be dimensionless, 
measured in units of the lattice spacing $a$.

The magnetic frustration $f$ is usually defined as the number of flux 
quanta per plaquette. 
There has been intensive research interest in the role of the magnetic frustration 
in both classical~\cite{ref:class} and quantum~\cite{ref:quantum} systems.
In the absence of positional disorder,
every plaquette has an equal area, thus $f$ is constant
over the whole system. 
It is well known that 
the Hamiltonian~(\ref{eq:H}) in this case is invariant under the transformation
$f \rightarrow f \pm 1$.~\cite{ref:class} 
In the present work, however, the plaquette area changes from place
to place, and the Hamiltonian loses the above symmetry, invalidating
the equivalence  between the cases $f$ and $f \pm 1$. 
We in this work define $f$ as the {\it disorder average}
(denoted to be $[ \cdots ]$) of the flux through one plaquette 
\begin{equation}
\left [ \sum_{p} A_{ij} \right ]  = 2\pi \frac{B a^{2} }{\Phi_{0}} 
\equiv  2 \pi f. \label{frustration}
\end{equation}  
The magnetic bond angle $A_{ij}$ in Eq.~(\ref{gauge}) is correlated with the nearest neighboring 
bond angles since a change of position $\mathbf{r}_i$ gives rise to changes of 
four magnetic bond angles $A_{ij}$ with site $j$ being neighbors of site $i$. 
Without such short-range correlations, $A_{ij}$ becomes a random quenched variable 
characterized by the disorder average 
$ \left [ A_{ij} \right ] = 2 \pi f \left( x_{j}^{0}+x_{i}^{0} \right)\left( y_{j}^{0}-y_{i}^{0} \right)$.  
In case that $f$ is an integer, one can gauge away the magnetic bond angle, 
establishing the equivalence with the RG$XY$ model
where the average frustration across the whole system vanishes. 
The variance of the sum of the magnetic bond angles around one plaquette in a PDJJA is given by
\begin{equation}
\label{SD_pdjja}
 \left[ \left( \sum_{p} A_{ij}  \right)^{2}  \right]  - \left[ \sum_{p} A_{ij} \right]^{2}
= \pi^{2} f^{2} \left( \frac{4}{3}\Delta^{2} + \frac{8}{9} \Delta^{4} \right)
\end{equation}
whereas the corresponding quantity in the RG$XY$ model reads
\begin{equation}
\label{SD_rgxy}
\left[ \left( \sum_{p} A_{ij}  \right)^{2}  \right] - \left[ \sum_{p} A_{ij} \right]^{2}
= \frac{4 \pi^{2}}{3} r^{2},
\end{equation}
where $A_{ij}$ has been taken to be uniformly distributed in $[-r\pi, r\pi]$ with the disorder strength $r$. 
It is to be noted that unless $\Delta$ is close to unity, one obtains 
\begin{equation} 
\label{eq:rDelta}
r \approx  f \Delta .
\end{equation} 
Consequently, the disorder strength of the PDJJA in comparison with 
the RG$XY$ model is measured by $f\Delta$, 
as found in previous studies of the PDJJA.~\cite{kosterlitz_pdjja, stroud}
It should be noted that the equivalence between the PDJJA and the RG$XY$
model becomes valid only when the external magnetic field yields
an integer value of the average magnetic frustration.  
The relation in Eq.~(\ref{eq:rDelta}) is practically very useful 
since it provides a convenient way to perform experiments on the 2D RG$XY$ model, 
in which the disorder strength can be tuned for one sample only by increasing the integral value of $f$.

We below sketch briefly the numerical method adopted in this work. 
Introducing the twist variable $\mathbf{D} \equiv \left( D_{x}, D_{y} \right)$ for the
fluctuating twist boundary conditions (FTBC)~\cite{kim2} in $L \times L$ arrays,
we write $\phi_{ij}$ in Eq.~(\ref{eq:H}) as
\begin{equation}
\phi_{ij} = \theta_{i} -\theta_{j} - \mathbf{r}^{0}_{ij} \cdot \mathbf{D}, 
\end{equation} 
where $\mathbf{r}^{0}_{ij} \equiv \mathbf{r}^{0}_{j} -\mathbf{r}^{0}_{i}$  
and $\theta_{i}$ is the phase angle.
Since the effects of positional disorder are introduced only through 
magnetic bond angles, we use the RSJ dynamics combined with the 
FTBC as follows: 
Using the local current conservation, the equation for the phase angle at site $i$ is given by
\begin{equation}
\dot{\theta}_{i} = - \sum_{j} G_{ij} {\sum_{k}}' 
 \left [\sin \left( \phi_{jk} -A_{jk} \right) + \eta_{jk} \right],
\label{eq:angle}
\end{equation} 
where time has been measured in units of $\hbar/2ei_c R$ with the single-junction 
critical current $i_c =2eE_J/\hbar$ and the shunt resistance $R$ 
(see Ref.~\onlinecite{kim2} for details), $G_{ij}$ is the square-lattice 
Green function, ${\sum_{k}}'$ denotes the summation over the four 
nearest neighbor sites of $j$ , and $\eta_{jk}$ is the dimensionless 
thermal noise current satisfying $ \langle \eta_{ij} \rangle = 0$ and
\begin{equation}
\langle \eta_{ij}(t) \eta_{kl}(0) \rangle = 2T \left(
\delta_{ik}\delta_{jl} - \delta_{il}\delta_{jk} \right) \delta(t),
\end{equation} 
with the temperature $T$ in units of $E_J/k_{B}$. 
In order to use the efficient fast Fourier transform, 
we modify the periodic boundary condition for $\theta_i$ according to 
\begin{eqnarray}
\theta_{i+L \hat{\bf x} } &=& \theta_{i} + 2 \pi f L \delta y_{i} \nonumber\\
\theta_{i+L \hat{\bf y} } &=& \theta_{i},  
\end{eqnarray}  
where $ 2 \pi f L \delta y_{i}$ is fixed in time and thus
$\dot{\theta}_{i + N \hat{\bf x} } = \dot{\theta}_{i}$. 
Here the usual periodic boundary conditions 
($\theta_{i + L\hat{\bf x}} = \theta_{j + L\hat{\bf y}} = \theta_{i}$)
are not applicable since $A_{ij} \neq A_{i{+}L\hat{\bf x}  ~  j{+}L\hat{\bf x}}$ for
the PDJJA.

We consider the system under uniform external currents injected along the $x$-direction. 
The Josephson relation $2eV_{x}/\hbar = \dot{\phi}_{i{+}N\hat{\bf x} ~   i}$, 
where $V_{x}$ is the voltage drop across the sample, together with the
global current conservation condition,~\cite{kim2} 
leads to the equations of motion for the twist variables $D_{x}$ and $D_{y}$ in the form
\begin{eqnarray}
\dot{D}_{x} & =  &\frac{1}{L^2} \sum_{{\langle ij \rangle}_{x}} \sin
\left( \phi_{ij} -A_{ij} \right) + \eta_{D_{x}} -i_{d}, 
\nonumber \\
\dot{D}_{y} & = &\frac{1}{L^2} \sum_{{\langle ij \rangle}_{y}} \sin
\left( \phi_{ij} -A_{ij} \right) + \eta_{D_{y}}, \label{eq:Dy}
\end{eqnarray} 
where $\sum_{{\langle ij \rangle}_{x(y)}}$ denotes the directed sum over 
nearest neighboring bonds in the $x(y)$ direction, $i_{d}$ is the external (driving)
current density in units of the single-junction critical current $i_c$, 
and $\eta_{D_{x}}$ ($\eta_{D_{y}}$) is the thermal noise current for
$D_{x}$ ($D_{y}$) satisfying 
$\langle \eta_{D_{x}} \rangle  = \langle \eta_{D_{y}} \rangle =0$ and
\begin{equation}
\langle \eta_{D_{x}}(t)\eta_{D_{x}}(0) \rangle = 
\langle \eta_{D_{y}}(t)\eta_{D_{y}}(0) \rangle = 2(T/L^2)\delta(t).
\end{equation} 

We numerically integrate the equations of motion given by Eqs.~(\ref{eq:angle}) 
and (\ref{eq:Dy}), and compute the $IV$ characteristics with the average (dc) voltage 
$V  \equiv  L v \equiv - L \langle \dot{D}_{x} \rangle$, 
where $v$ denotes the voltage drop per junction in units of $i_c R$. 

\section{\label{sec:result1} Results for weak disorder}


Our simulation scheme requires sufficiently precise data for a large sample
and a large disorder strength.  We find that RSJ simulations of the PDJJA on 
a $128\times 128$ square lattice can be made to meet all these three
requirements simultaneously, and thus use this lattice size in the present
investigation.

To probe phase transitions in weak and strong disorder regimes, 
we compute the $IV$ characteristics at various temperatures and values of $f$ 
with the latter controlling the disorder strength (see Sec.~\ref{sec:model}). 
In our simulations we set the parameter for positional disorder 
equal to $\Delta = 0.2$.
Since $r \approx f \Delta$ and the critical disorder strength in the 
RG$XY$ model is believed to be $r_{c} \approx 0.4$,~\cite{kosterlitz,akino,holme}
we vary the average frustration $f$ from $1$ to $4$ 
to cover both weakly ($f=1$ with  $f \Delta \approx 0.2 <r_{c}$)
and strongly ($f\geq 2$ with $f \Delta  > r_{c}$)
disordered cases.

\begin{figure}
\includegraphics[width=0.45\textwidth]{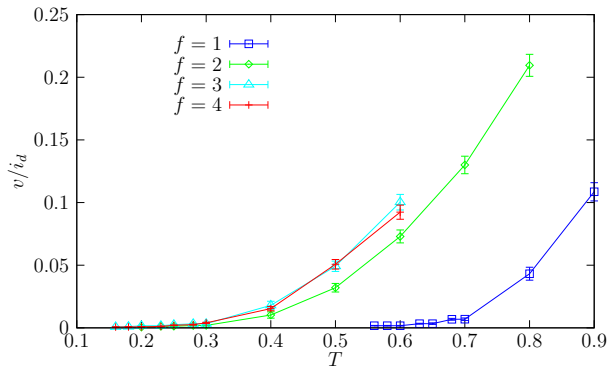}
\caption{\label{RT} (Color online) Resistance $v/i_d$ (in units of $R$) versus 
temperature $T$ (in units of $E_J/k_{B}$) for various frustrations 
in positionally disordered Josephson-junction arrays. 
As the frustration is raised from $f=1$ to $f=2$, 
the effective disorder strength becomes larger, reducing the critical temperature. 
Larger values of $f \, (=3,4)$ yield only insignificant variations of 
the critical temperature.
}
\end{figure}

Figure~\ref{RT} presents the dc resistance $v/i_d$ (in units of the shunt resistance $R$)
versus the temperature $T$ at various frustrations $f=1, 2, 3$, and $4$. 
When $f=1$, corresponding to the weakly disordered case, 
the resistance-temperature (RT) curves in Fig.~\ref{RT} 
disclose that the system undergoes a phase transition around $T=0.6$, which is
lower than $T_{c}=0.89$ in a regular JJA.~\cite{kim2}
For strong disorder ($f \geq 2$), phase transitions are observed to
occur near $T=0.2$, hardly depending on the frustration.      

In order to determine the transition temperature in a more accurate 
way and to understand the dynamic critical behavior in detail, 
we next investigate the $IV$ characteristics at various values of $f$ 
with the help of the scaling form suggested in Ref.~\onlinecite{fisher}. 
In two dimensions, the current density and the electric field behave as 
$J \sim T /\xi$ and $E \sim \xi^{-1-z}$, respectively, with the correlation length $\xi$.
In the case $T_{c} \neq 0$, one replaces $T$ in $J \sim T /\xi$ by $T_{c}$ near
criticality, and obtains accordingly $J \sim \xi^{-1}$. 
In this work, $i_d$ and $v$ 
correspond to the current density and the electric field, respectively, 
and yield the scaling form
\begin{equation}
v=i_{d} \xi^{-z} F_{\pm}\left( i_{d} \xi \right)
\label{scaling}
\end{equation} 
with the scaling function $F_{\pm}$ above/below $T_{c}$.
At $T=T_{c}$, the correlation length $\xi$ diverges and Eq.~(\ref{scaling}) leads to
$v \approx i_{d}^{z+1}$.

\begin{figure}
\includegraphics[width=0.45\textwidth]{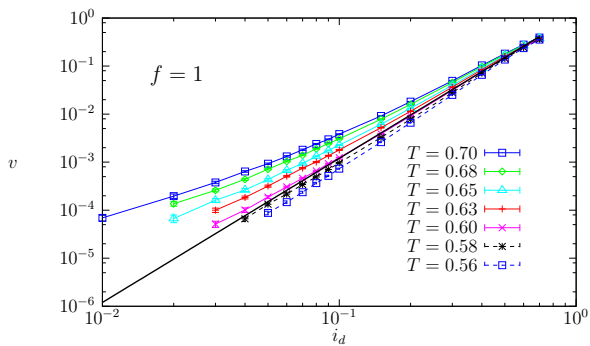}
\caption{\label{IV_f1}(Color online)  $IV$ curves ($v$ versus $i_{d}$) for $f=1$ 
in log scales at various temperatures $T$ (in units of $E_J/k_{B}$). 
At $T \lesssim 0.6$, curves fit well to the power-law form
$v \approx i_{d}^{z+1}$, with the solid line representing the case $z=2$. 
At higher temperatures, $IV$ curves bend upwards, indicating a BKT-type transition.
}
\end{figure}

\begin{figure}
\includegraphics[width=0.45\textwidth]{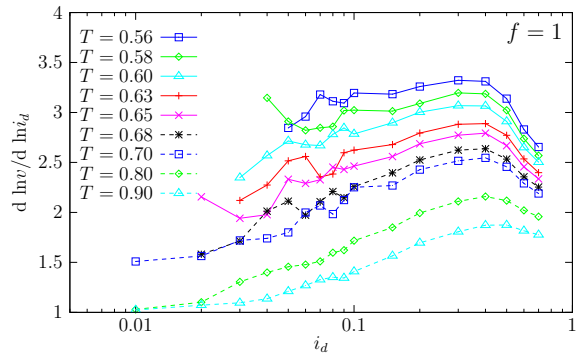}
\caption{\label{diff_f1}(Color online)  
Slope $d\ln v /d\ln i_{d}$ of the $IV$ curve 
versus the driving current $i_{d}$ at various temperatures (in units of $E_J/k_{B}$). 
Due to finite-size effects, the slope first increases with $i_{d}$, 
then decreases toward the Ohmic value, unity, thus forming a peak. 
In a broad range of temperatures below $T_c \approx 0.7$, the peak position
$i_{d}^{\ast}$ does not change substantially while it shifts to higher currents
beyond $T_c$.  Such behavior of the peak position is consistent with the BKT-type transition. 
}
\end{figure}

In Fig.~\ref{IV_f1}, we display the $IV$ characteristics for $f=1$,
corresponding to the weak disorder regime, 
near $T = T_c \approx 0.6$ (see the RT curve for $f=1$ in Fig.~\ref{RT}). 
Due to the finite-size effects, $IV$ curves exhibit the Ohmic behavior
in the small current region as $i_{d} \rightarrow 0$.~\cite{kim2} 
The curves at $T \lesssim 0.6$ fit well to the power-law form 
$v \sim i_d^a$: At $T \approx 0.6$,  we have $a \approx 3$, and 
the lower $T$, the larger $a$. In contrast, the $IV$ curves
at higher temperatures ($T \gtrsim 0.6$) clearly exhibit upward
curvature, indicating the existence of a finite current scale (and thus of a finite length scale). 
From the relation $a = z+1$ between the dynamic
critical exponent $z$ and the nonlinear $IV$ exponent $a$, we thus reach
the conclusion that a phase transition of the BKT nature occurs 
at $T_{c} \approx 0.6$ with $z \approx 2$. 
It should be noted that the scaling form in Eq.~(\ref{scaling}) is valid
when the correlation length diverges only at $T_c$ and becomes smaller
as the temperature is varied (either lowered or raised) from $T_c$. 
Accordingly, when the phase transition is of the BKT type, characterized
by the diverging length scale in the whole low-temperature phase, 
only high-temperature $IV$ curves collapse to the scaling form, while 
the low-temperature part cannot be made to collapse due to the lack
of the length scale $\xi$.

Alternatively, one can also plot the slopes of $IV$ curves, given by 
$d\ln v /d\ln i_{d}$, versus the driving current $i_{d}$, 
as shown in Fig.~\ref{diff_f1} to confirm the BKT nature 
of the transition.~\cite{medvedyeva} 
As $i_d$ is reduced, all curves should eventually crossover to the Ohmic
behavior characterized by the unit slope ($d\ln v /d\ln i_{d} = 1$) 
due to the finite-size effects. 
The peak position $i_d^\ast$ of each curve in Fig.~\ref{diff_f1} 
measures a characteristic current scale, which is inversely proportional to the length
scale in the system, i.e., $i_{d}^{\ast} \approx i/\xi$.
The correlation length $\xi$ increases as $T$ approaches $T_{c}$ 
from high temperatures. Near $T_c$ in the high-temperature phase
and also in the whole low-temperature phase, the correlation length
becomes larger than the size of the system. In this case, the relevant
length scale of the system is not the correlation length but the
linear size $L$ of the system, leading to $i_{d}^{\ast} \sim 1/L$ 
independent of the temperature. 
In Fig.~\ref{diff_f1}, the peak position $i_{d}^{\ast}$ has almost the same value
around $0.4$ as $T$ is increased from below up to $T \approx 0.7$; then
$i_{d}^{\ast}$ appears to drift away toward larger values as $T$ is increased
further beyond $T=0.7$, which is in good agreement with what we expect
for the BKT-type phase transition. We emphasize again that
for a finite system $i_d^\ast$ starts to move not at
the true critical temperature but at a temperature higher than $T_c$ when
the correlation length becomes comparable to $L$. 
Between $T=0.63$ and 0.6, $d\ln v /d\ln i_{d} \approx 3$ at $i_{d}^{\ast}$,
which, combined with the power-law decay form in Fig.~\ref{IV_f1}, 
makes us conclude that the BKT transition occurs at 
$T_c \approx 0.6$ (in accord with the RT curve in Fig.~\ref{RT}) 
with the dynamic critical exponent $z \approx 2$.
This analysis of the $IV$ curves clearly reveals that the phase transition of 
the BKT type occurs in the weak positional disorder regime of the PDJJA, 
consistent with experiment~\cite{yun} and previous numerical studies of RG$XY$ model.~\cite{kosterlitz,akino, holme}

\section{\label{sec:result2} Results for strong disorder}

\begin{figure*}
\centerline{\includegraphics[width=0.9\textwidth]{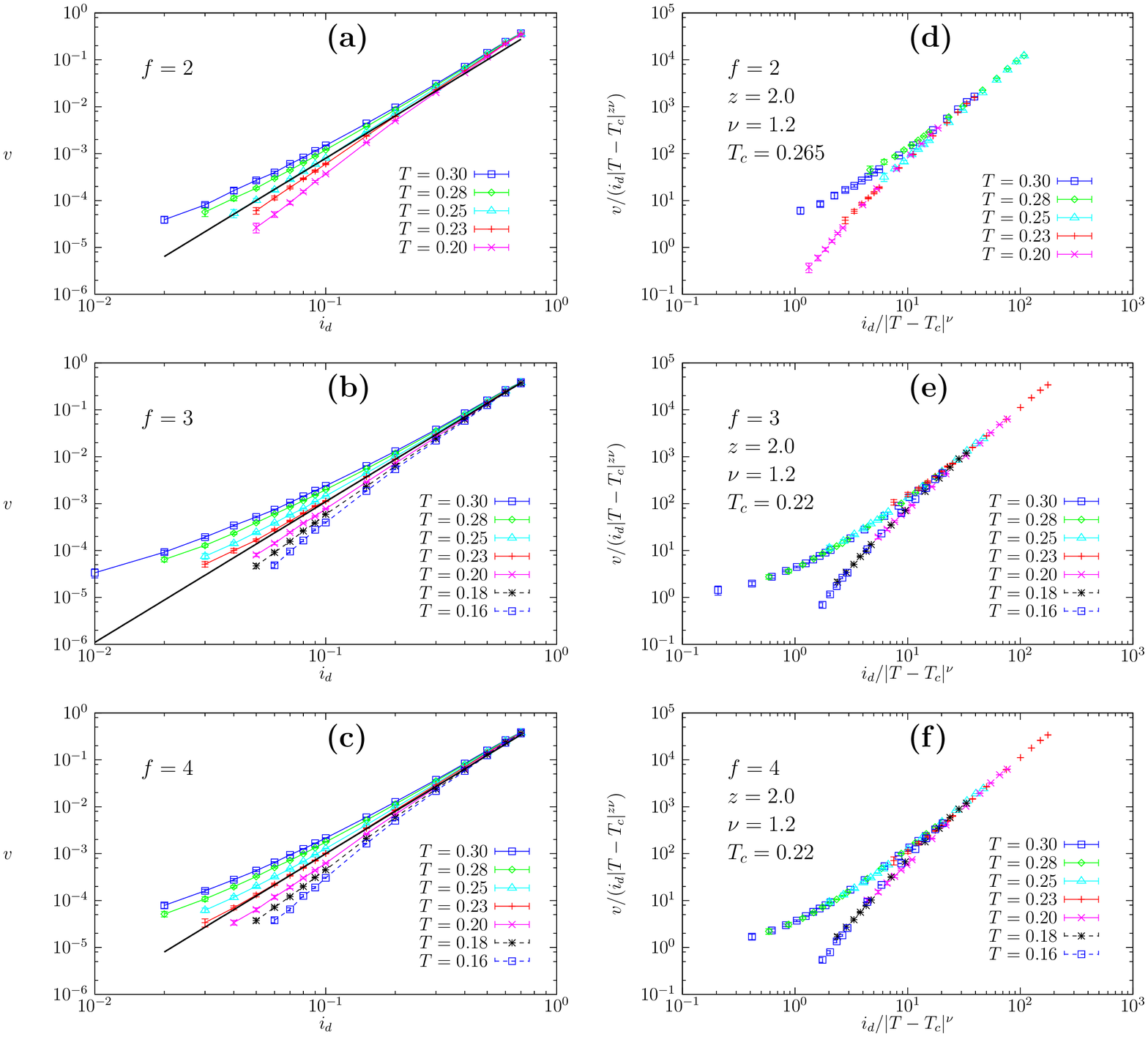}} 
\caption{\label{fig:strong}(Color online)  
$IV$ characteristics ($v$ versus $i_{d}$) at various temperatures 
for $f =$ (a) $2$, (b) $3$, and (c) $4$, and the corresponding
scaling plots in (d)-(f). 
The solid lines in (a), (b), and (c) represent the power-law decay form 
$v \approx i_{d}^{z+1}$ with $z=2$. 
A well-defined temperature $T_c$ separates the $IV$ curves 
into two groups, one bending upwards and the other downwards.
All data points in the corresponding $IV$ characteristics 
are made to collapse into scaling functions, as shown in (d)-(f), with 
$T_{c} = 0.265, 0.22$, and $0.22$, respectively. 
In all the three cases (d)-(f), the same critical exponents $\nu =1.2$ and $z=2.0$ 
are used, implying that the transitions in the strong disorder regime ($f=2, 3$, and $4$) 
belong to the same universality class.}
\end{figure*}

In the strong disorder regime ($f \geq 2$), we show in Fig.~\ref{fig:strong} 
the $IV$ curves for $f =$ (a) $2$, (b) $3$, and (c) $4$. 
It is clearly observed that for each value of $f$,
there exists a well-defined temperature $T_c$ at which the $IV$ curve
shows a power-law behavior, manifesting the absence of a length
scale at criticality. 
The critical temperature estimated from Fig.~\ref{fig:strong} is
$T_c =$ (a) $0.25$, (b) $0.23$, and (c) $0.23$ for $f= 2, 3$, and $4$, 
respectively, in agreement with Fig.~\ref{RT}. It is also shown 
in Fig.~\ref{fig:strong} that the $IV$ curves at $T_{c}$ fit 
well to the form $ i_{d}^{z+1}$ with the dynamic critical exponent $z=2$. 
The $IV$ curves in Figs.~\ref{fig:strong} (a), (b), and (c) all
bend upward above $T_c$, indicating that the PDJJA is in the
high-temperature normal phase, while the opposite downward curvature 
below $T_c$ implies the superconducting phase in the limit of 
$i_d \rightarrow 0$.

To find nature of the phase transition and the critical temperature 
together with critical exponents in the strong disorder regime of the PDJJA, 
we employ the scaling form in Eq.~(\ref{scaling}),
which, with the correlation length $\xi \sim |T-T_{c}|^{-\nu}$, 
reads
\begin{equation}
\frac{v}{i_{d} |T-T_{c}|^{z\nu}} = F_{\pm} \left(
i_{d} |T-T_{c}|^{-\nu} \right).
\label{scaling2}
\end{equation} 
Figures~\ref{fig:strong} (d)-(f) show that data points in the $IV$ characteristics 
[in Figs.~\ref{fig:strong} (a)-(c)] collapse into two function $F_{\pm}$ in Eq.~(\ref{scaling2});
this confirms that for given value of $f$, there exists a finite-temperature phase 
transition of a non-BKT type. 
The critical temperature changes from $T_c \approx 0.265$ for $f=2$ to 
$T_c \approx 0.22$ for $f=3$ and $4$, which is consistent with the experiment on 
the PDJJA~\cite{yun} and numerical studies of the RG$XY$ model~\cite{holme} 
and the gauge-glass model.~\cite{li,choi,kim,chen,holme2} 
On the other hand, in all three cases [$f=2, 3$, and $4$ in Figs.~\ref{fig:strong} (d)-(f)], 
the scaling collapse is achieved with $\nu = 1.2$ and $z = 2.0$, implying
that the nature of the transition in the strong disorder regime remains unchanged
as the disorder strength is increased.  
Furthermore, the dynamic critical exponent $z=2.0$ agrees with the studies of dynamic
behavior in the gauge-glass model~\cite{choi,kim,chen} as well as with the
experimental results~\cite{yun}. 
The critical exponent  $\nu =1.2$ obtained in this work is again 
consistent with the previous numerical results~\cite{holme,choi,kim,chen}
but not with the experimental result for the PDJJA,~\cite{yun} 
the origin of which is not clear at this stage. 

\begin{figure}
\includegraphics[width=0.45\textwidth]{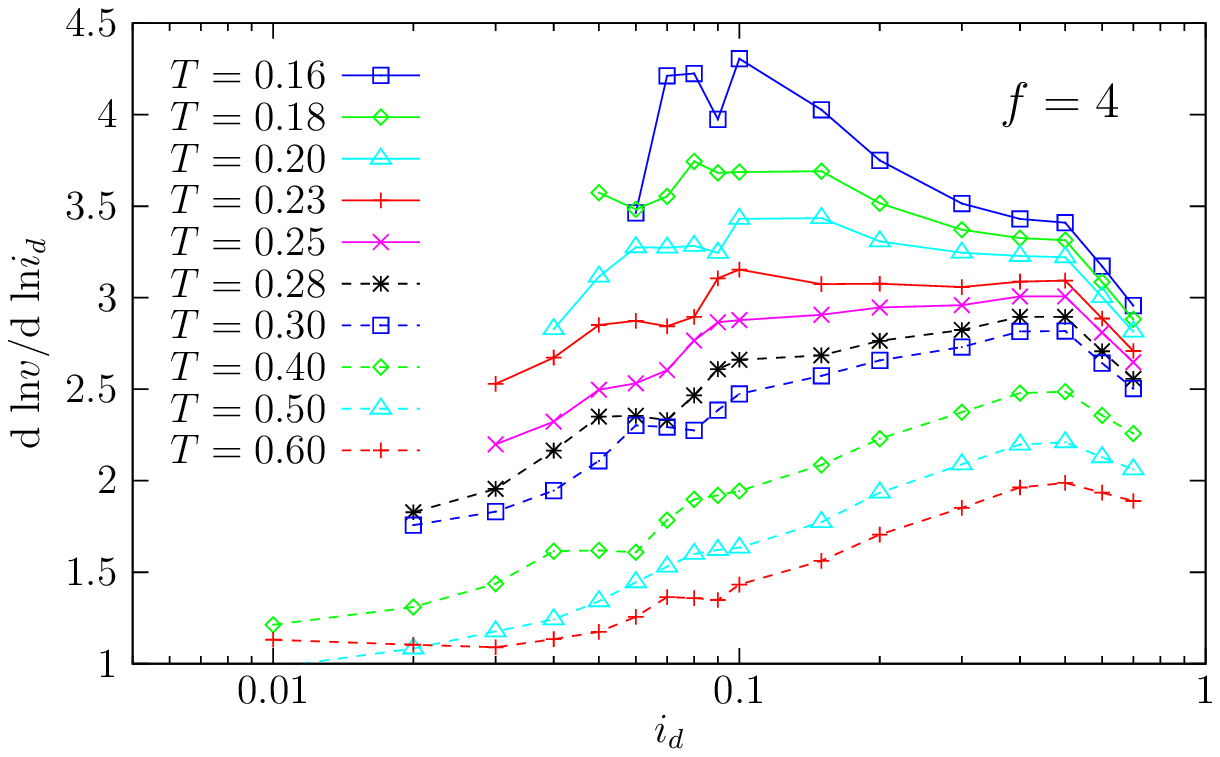}
\caption{\label{diff_f4}(Color online)  
Slope ($d\ln v/ d\ln i_{d}$) of $IV$ curves against $i_{d}$ for $f=4$,
corresponding the strong disorder regime of the PDJJA, 
at various temperatures $T$ (in units of $E_J/k_{B}$). 
Power-law behavior appears at $T=0.23$ to $0.20$ with the dynamic 
critical exponent $z \approx 2$, indicating the existence
of a phase transition with a diverging length scale only at $T_c$. 
}
\end{figure}

We next investigate slopes of the $IV$ curves for $f=4$ in the
same manner as in Sec.~\ref{sec:result1} for the weak disorder case for $f=1$. 
In Fig.~\ref{diff_f4}, the slope $d\ln v/d\ln i_{d}$ is plotted as a function of $i_{d}$ at
various temperatures around $T_{c}$.  Note that finite-size effects yield
the Ohmic behavior, $d\ln v/d\ln i_{d} \approx 1$, as $i_{d} \rightarrow 0$. 
It is observed that the curves at $T=0.23$ and $0.20$ are almost horizontal 
in broad ranges of the external currents, indicating scale-free behavior 
and in turn the existence of a phase transition 
with the dynamic critical exponent $z \approx 2.0$, 
consistent with the finding in Fig.~\ref{fig:strong}(f) for $f=4$.
Furthermore, the fact that the peak position changes rather abruptly 
from large currents to smaller currents at $T=0.23$ to $0.20$ implies 
that the phase transition is of a non-BKT type, namely, 
the $IV$ data are well scaled to two different functions, $F_{\pm}$ in Eq.~(\ref{scaling2}) 
above/below $T_{c}$.  

In view of the ongoing controversy about the existence of a finite-temperature
transition of a non-BKT character for strong disorder, one might ask what our
results really imply.  First of all, our evidence has been obtained for a
finite sample with one disorder realization (albeit a very large one in
comparison with those in most of earlier investigations).  Does the phase
transition survive in the large system size limit?  Since a single positional
disorder for a finite system can be periodically repeated, we can obtain an
arbitrarily large system by just adding new squares with the same single
disorder realization. Such an infinite system constructed from a single
disorder realization will, to our belief, most certainly have a
finite-temperature transition of a non-BKT type. This means that the phase
transition does exist {\it per se}.  Suppose that we instead choose an
arbitrarily large system and generate the disorder in the same random way as we
have done for our $128\times 128$ sample. Would the phase transition still
survive?  Here we find that various randomly generated disorder realizations
for the $128 \times 128$ system yield numerically very similar results,
indicating that for the size $L=128$ there is already a large amount of
disorder self-averaging. This again suggests that our results will survive in
the large $L$ limit, i.e., for the PDJJA with uniformly distributed disorder,
chosen randomly, of strength $\Delta=0.2$. In order to check the
finite-size effect on the self-averaging property in a more careful way, 
we have also computed voltages at $T=0.16$ in the strong disorder regime
($f=4$) for smaller 
sizes ($L=16, 32, 64$, and 128) at two different disorder realizations : 
The difference between voltages obtained 
from different disorder realizations is found to decrease as $L$ is increased, 
and becomes negligibly small at $L=128$. This indicates that indeed
the self-averaging effect becomes clear beyond $L=128$.
The next question is then whether
our results also carry over to the RG$XY$ model and the gauge-glass model.
Here the evidence is more circumstantial and based, on the one hand, on the
strong connection between the PDJJA and the RG$XY$ and random gauge-glass
models~\cite{forrester,kosterlitz} and, on the other hand, on the strong
similarity of the present results to some earlier results obtained from the
latter models.~\cite{holme,choi,kim,chen}

\section{\label{sec:summary}
Summary}
The existence of a finite-temperature phase transition 
in the strong disorder regime of the PDJJA and the 2D RG$XY$ model, including
the fully disordered case of the 2D gauge-glass model, are still in an
intensive debate. This work has been motivated by the very recent experiment
on $800 \times 200$ Josephson junction arrays with positional disorder,~\cite{yun}
and explored numerically the dynamic critical behavior of the PDJJA 
in transverse magnetic fields. 
Adopting the RSJ dynamics,~\cite{kim2} we have computed the $IV$ characteristics
of the PDJJA with the positional disorder parameter $\Delta =0.2$ for 
frustration $f=1, 2, 3$ and $4$.
The relation $r \approx f \Delta$ between 
the disorder strength $r$ in the 2D RG$XY$ model and the parameter $\Delta$ 
in the PDJJA, combined with the critical disorder
strength $r_{c} \approx 0.4$,~\cite{kosterlitz, akino, holme}
implies that the PDJJA for $f=1$ corresponds to the weakly disordered case
while strong frustration ($f \gtrsim 2$) puts the PDJJA in the strong disorder regime. 
The scaling analysis~\cite{fisher} of $IV$ curves and their slopes~\cite{medvedyeva} 
has revealed clear evidence for $T_c \neq 0$ in the PDJJA with strong disorder, 
which agrees with the experiment on the PDJJA~\cite{yun} as well as previous numerical studies
of the RG$XY$ model~\cite{holme} and the gauge-glass model.~\cite{li,choi,kim,chen,holme2}

\begin{figure}
\includegraphics[width=0.45\textwidth]{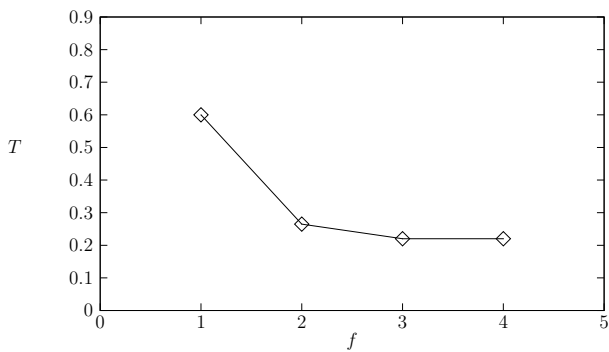}
\caption{\label{phase} 
Phase diagram on the plane of temperature $T$ (in units of $E_J/k_{B}$) 
and frustration $f$ (controlling the disorder strength). 
The solid line, separating the high-temperature normal phase from
the low-temperature superconducting phase, is only a guide to the eye.
The critical temperature saturates to a finite nonzero value
$T_c \approx 0.22$ as $f$ is increased.
Below $T_{c}$, depending on the disorder strength, there exist two different 
superconducting phase: a BKT-like phase for $f=1$
and a non-BKT type one for $f \gtrsim 2$.
}
\end{figure}

Figure~\ref{phase} exhibiting the phase diagram on the plane of the
temperature and frustration summarizes the results of the present work. 
The critical temperature $T_{c}$ is observed to reduce rapidly 
as the average frustration $f$, controlling the disorder strength, is increased 
in the weak disorder region. 
It then appears to saturate toward a constant value $T_{c}=0.22$ 
in the strong disorder regime.
Below the phase boundary separating the superconducting phase at low
temperatures and the normal phase at high temperature, 
there exist two different superconducting orders, according to the disorder strength: 
In the weak disorder regime, e.g., $f=1$, the superconducting state
is the low-temperature BKT phase characterized by the divergence
of the correlation length.  On the other hand, for strong disorder ($f \gtrsim 2$), 
the transition is of the non-BKT type with the well-defined correlation length 
critical exponent $\nu =1.2$, 
consistent with the value obtained previously for the 2D gauge-glass 
model~\cite{choi, kim, chen} but inconsistent with the experimental
finding $\nu = 2.0 \pm 0.3$.~\cite{yun}
The origin of this discrepancy is not clear at present and needs more detailed investigation in the future study. 
Finally, we point out that the resemblance between the phase diagram in Fig.~\ref{phase}
and the corresponding diagram obtained in the recent experiment on the PDJJA~\cite{yun}
is striking, which is also in very good agreement with 
the numerical study of the 2D random gauge $XY$ model.~\cite{holme}

\acknowledgments

This work was supported by the Korea Research Foundation Grant funded
by the Korean Government (MOEHRD) KRF-2005-005-J11903 (B.J.K.),
through the National Research Center for Systems Bio-Dynamics (M.Y.C.), 
and through the Creative Research Initiatives Program (S.-I.L.).


\bibliography{apssamp}

\begin{thebibliography}{}
\bibitem{footnote} 
We assume that a change of the central position of a superconducting island in
a Josephson junction array does not alter the Josephson coupling strength, 
which is related with the tunneling amplitude of Cooper pairs 
(see Ref.~\onlinecite{yun} for more details).
\end{thebibliography}

\end{document}